\documentclass[aps]{revtex4}
\usepackage{graphics}

\begin{document}

\newcommand{\gb} {{\bf g}}
\newcommand{\gbt} {\tilde{{\bf g}}}
\newcommand{\gm} {{\mathrm {g}}}
\newcommand{\gmt} {\tilde {mathrm {g}}}

\title{\normalsize{\bf{COHERENT BREMSSTRAHLUNG  IN IMPERFECT PERIODIC
 ATOMIC STRUCTURES}}}

\author{ S. Bellucci$^1$, V.A. Maisheev$^2$\\
 $^{1}$\it{ INFN - Laboratori Nazionali di Frascati, P.O. Box 13, 00044 Frascati, Italy} \\
 $^{2}$\it{ Institute for High Energy Physics, 142281, Protvino, Russia }}

\begin{abstract}
Coherent bremsstrahlung of high
energy electrons  moving in a three-dimensional imperfect
periodic lattice consisting of a complicated system of atoms
is considered. On the basis of the normalized probability
density function of the distribution of atomic centers
in the fundamental cell the relations describing coherent and
incoherent contributions into  cross sections are obtained.
In particular, the cross section of coherent bremsstrahlung
in complex polyatomic single crystals is found.
 The peculiarities of formation and possibilities of
utilization of coherent processes are discussed.

\end{abstract}
\maketitle
\section{Introduction}
For the first time, the theory of the coherent bremsstrahlung and
electron-positron pair production in single crystals
was published in Ref. \cite{Fr,TM,U}.
Up to now, these processes have been much studied both
theoretically \cite{Tim,HB,BKS}  and experimentally \cite{Barb,Timm,CB}.
The specific peculiarities of the coherent processes were widely used
for obtaining linearly polarized $\gamma$-beams in polarization
measurements \cite{MV,Bas,UE} and for increasing the $e^{\pm}$-beam intensity \cite{AMF}.
These experimental investigations were carried out in a wide energy
range of electron, positron and $\gamma$-beams from a few  hundreds of MeV to
 100 - 200 GeV. As a rule, simple crystallographic structures were
used in  experiments. Both experiments and theory show that, with
the increasing of the particle energy, the requirements on the beam divergence
become more strict. Besides, at high enough energies the process of coherent
bremsstrahlung is violated, due to the magneto-bremsstrahlung mechanism \cite{BKS}.
This mechanism provides the linear polarization of the emitted
$\gamma$-quanta. However, a sufficiently small angle divergence of the electron
beam is needed for utilizing this phenomenon.
One of the main requirements, which restricts the use of coherent
processes, is the small size of the fundamental cell of the single crystals, which has to
be of the order of some
angstroms. It is our opinion that,
finding atomic periodic structures with a minimal period of
tens or hundreds angstroms, may yield a good solution to the above-mentioned problem.

In recent years, considerable advances
have been made in the creation of various nanostructures \cite{AT, Dress, KK},
such as  regular two-dimensional arrays, fullerite crystals,
nanofilms, nanotube superlattices and so on. In a number of papers \cite{KL,GII,ZG}
 nanotube lattices were considered as a source of channeling radiation.
Other applications of nanotubes for purposes of high energy physics
are also described in Ref. \cite{BM,BB,BB1}.

According to Ref. \cite{AT} single-wall nanotubes are uniform in diameter
and self-organized into ropes, which consist of 100 to 500 nanotubes
in a two-dimensional triangular lattice with a lattice constant of 17
angstroms.
In our talk \cite{BM} we discussed the processes of
coherent bremsstrahlung  and $e^\pm$-pair production in the nanotube
superlattice. In considering this task we  met  problems, the solution
of which has a common meaning for the above-mentioned coherent processes in various
nanostructures and complex single crystals. Below we give an introduction to
these problems, using the example of the nanotube superlattice.

 Fig. 1  illustrates the three-dimensional superlattice of (10,10) armchair
single wall nanotubes. In this case we can write for the vector-radius
${\bf{r}}_j(x_j,y_j,z_j)$ of the  $j$-th  atom in the nanotube
\begin{equation}
\label{1}
x_{1,j}=R\cos{({{4\pi j}\over {N}} + \varphi_1)},\,\,
x_{2,j}=R\cos{({{4\pi j}\over {N}}+\varphi_2)},
\end{equation}
\begin{equation}
\label{2}
y_{1,j}=R\sin{({{4\pi j} \over {N}} +\varphi_1)},\,\,
y_{2.j}=R\sin{({4{\pi j} \over {N}}+ \varphi_2)},
\end{equation}
\begin{equation}
\label{3}
 z_{1,j}=0, \,\, z_{2,j}=b/2,
\end{equation}
where $j\, = 1,2,...,N/2 $ are the indices corresponding to atoms
placed
in two parallel planes, $R$ is the radius of the ring, $b$ is the
period
 (i.e. the size of
the fundamental cell) in the $z$-direction, $\varphi_1,\, \varphi_2$ are the angle
 shifts ($\varphi_1-\varphi_2=$ const).

In Fig. 1 we describe an ideal nanostructure where  the angle shifts of all
nanotubes are  the same. The experiments \cite{AT} show that these
angle shifts are distributed randomly or, in other words, every nanotube
 is turned at some different angle.
 This means that the content of every
cell will be different, relative to the coordinate system. Thus,
the nanotube lattice is not a periodic structure in a strict sense,
in spite of the  constancy of the distance between
neighboring nanotubes. However, the existing theory of coherent
bremsstrahlung holds its validity for atomic structures (single crystals),
which are periodic in a strict sense.

Based on this example, we formulate the common problem for the calculation
of the coherent bremsstrahlung in artificial and natural nanostructures.
The problem is the violation of the periodicity in a strict sense
in these structures.

This situation is well known in the diffraction physics of x-rays \cite{Cow_A} in
imperfect structures. However, the process of coherent bremsstrahlung was
investigated and utilized mainly in simple single crystals. Such crystals as
silicon and diamond have a negligibly small degree of mosaicity and admixture.
Because of this fact, the problem of calculation of the bremsstrahlung in  imperfect
structures did not practically appear (except for the problem of the thermal atomic
motion \cite{TM}).
In the x-ray diffraction theory the above-mentioned problem was solved with the
help of the introduction of the averaged electron density \cite{Cow_A}. In our talk \cite{BM},
using the analogy in the description of the diffraction and coherent processes,
we could solve this problem on the basis of physical sense. Besides, we
suggest another approach which is based  on computer simulations. We think that
this approach may be extended to a wide class of analogous problems,
in particular, for a nanotube lattice with a more complicated dependence
on the angle shifts, than a random one.

It is turns that our computer approach has an analytical solution in the general case.
Furthermore, on the basis of our method we are able to consider the process of
coherent bremsstrahlung in imperfect atomic structures, taking into account all fluctuations factors.

The paper is organized as follows. In section II we give
a mathematical introduction in the problem.
In section III, for the description of fluctuations in atomic structures,
we introduce the normalized probability density function and formulate
some rules for averaging of the structure factors. In section IV we consider
the three dimensional model of the real atomic structure with fluctuations.
The results obtained here allow us to derive (in section V) the coherent and
incoherent cross sections of the bremsstrahlung process in imperfect structures.
In section VI we discuss the influence of thermal fluctuations
in atomic structures on the coherent bremsstrahlung. Here we reproduce
well known results and also obtain new ones. In section VII we consider
the possibility of generalizing our theory to consider multiatomic structures and,
in particular, multiatomic single crystals. Samples of calculations
of the bremsstrahlung process in real atomic structures
are presented in section VIII. In conclusion (in section IX) we give
shortly the main results of our investigations.

\section{Cross section of coherent bremsstrahlung in  ideal periodic
structures}

The differential cross section  of the coherent bremsstrahlung for
ideal periodic structure, consisting of atoms, can be written
in the following form \cite{TM}:
\begin{equation}
\label{4}
d\sigma_{CB}=d\sigma_{BG} |\sum_i e^{{i{\bf{q}} {\bf{r}}_{i0}/\hbar}}|^2,
\end{equation}
where
$\sigma_{BG}$ is the bremsstrahlung cross section for an isolated
atom, ${\bf{q}}$ is
the three-dimensional transfer momentum,
${\bf{r}}_{i0}$ are the vector-radii
of the atoms in the periodic structure.

From this expression, the following relation for the cross section
 per atom \cite{TM} is derived:
\begin{equation}
\label{5}
d\sigma_{CB}(E,E_{\gamma}, {\bf{q}})={(2\pi)^3 \over {N V}}
\sum_{\gb}d\sigma_{BG}(E,E_{\gamma}, {\bf{q}}) |S(\gb)|^2
\delta({\bf{q}}/\hbar -\gb),
\end{equation}
where $N$ is the number of atoms in the fundamental cell of the
structure, $V$ is the volume
of the fundamental cell,
$S(\gb)$ is the structure factor \cite{TM,Tim}, $\gb$ is the
vector of the reciprocal lattice, $\delta$ is the delta-function,
$E, E_{\gamma}$ are the energies of the initial electron and
bremsstrahlung $\gamma$-quantum and $\hbar$ is the Planck constant.
The structure factors are calculated from the relation
\begin{equation}
\label{6}
S(\gb)=\sum_{j=1}^{N}e^{{i\gb {\bf{r}}_j}},
\end{equation}
where ${\bf{r}}_j$ is the radius of the $j$-th atom in the
fundamental cell.

From Eq. ({5}) one can see that the specific character of every atomic
structure is defined by its structure factors.
It is obvious that for ideal structures (at a fixed localization of the atoms)
 the structure factors take well defined values. Because of various fluctuations,
the coordinates of atoms in the fundamental cell are changed with
space and (or) time and this fact does not allow to use Eq. ({5})
for calculations.
For this reason,
it is necessary to understand the behavior of the structure
factors for these fluctuations.

The plan of our further actions for solving the above-mentioned problem
is the following: we will try to reduce the problem
to one, for which the solution is known
(such as the process in the ideal periodic lattice).
In the first stage of the study, we will formulate the definition and some rules
for the averaged structure factors. Then, we will consider the simulations
of fluctuations on the model of real periodic structures, and thereafter
we will use the obtained results for the calculation of
the coherent bremsstrahlung cross
section  in  imperfect atomic structures.

It should be noted that in the theory of coherent bremsstrahlung
the potential of the crystal is considered as the sum
of isolated atomic potentials. It is obvious that this assumption is only
approximately true. However, the current experimental experience
(see, for example Ref. {\cite{Tim}})
shows the correctness of this statement with a high enough accuracy.
In this paper we will also hold this statement true and because of this,
our results will be easy to compare with the standard theory.
In the following, we will make use of the expression "coordinates of the atomic center"
which has an exact physical meaning denoting  the coordinates of the
atomic nucleus.

\section{Averaging of the structure factors}

As previously noted,  the specific character of
every structure is defined by its structure factors.
It is useful to appreciate the physical meaning of these quantities. For this purpose, we write
the atomic density for a periodic structure in the point given by
the vector-radius ${\bf{r}}$
\begin{equation}
\label{7}
n_a ({\bf{r}}) =\sum_{k} \sum_{j=1}^{N} \delta ({\bf{r}} -{\bf{r}}_k
-{\bf{r}}_j) = {N \over V} + {1\over V} \sum_{\gb} S(\gb)
e^{-i\gb {\bf{r}}}.
\end{equation}
From here, it follows that $S(\gb)/V$ is the Fourier component of the atomic density
or, in other words, the structure factors are the atomic images in the reciprocal space.
Note that Eq. (7) does not take into account thermal atomic fluctuations. They are easy
to calculate with the help of the following multiplier: $\exp(-A\gb^2/2)$
(see below).
We stress that the structure factors depend on the choice of a coordinate system
and therefore the values of structure factors have a physical meaning only in
a defined coordinate system.

The space distribution of the atomic centers in the fundamental cell
of the structure  can be described with the help of the
normalized probability density function
${\cal{P}}({\bf{x}}_1,{\bf{x}}_2,..., {\bf{x}}_{N})$,
where ${\bf{x}}_1,...,{\bf{x}}_{N}$ are the space displacements
of the atomic centers from the points ${\bf{r}}_j$ in the cell. The integral
of this function over the whole ($3 \times N$-dimensional)${\cal{V}}$ volume
of the cell is equal to 1. Then the structure factor, averaged
with the help of ${\cal{P}}$-function, is given by
\begin{eqnarray}
\label{8}
 \langle S(\gb) \rangle =  \int S(\gb,{\bf{r}}_1 - {\bf{x}}_1,..,
{\bf{r}_{N}}- {\bf{x}}_{N}) {\cal{P}}
({\bf{x}}_1,{\bf{x}}_2,..., {\bf{x}}_{N})
d{\cal{V}} = 
\sum^{N}_{j=1}
e^{i{\bf{r}}_j \gb}
\int_V e^{-i{\bf{x}}_j \gb}
{\cal{F}}_j({\bf{x}}_j)
d{\bf{x}}_j\, ,
\end{eqnarray}
where the ${\cal{F}}_j$-function reads
\begin{equation}
\label{9}
{\cal{F}}_j({\bf{x}}_j) =\int {\cal{P}} d{\cal{V}}(-j) \, ,
\end{equation}
with
$d{\cal{V}}(-j)=
d{\bf{x}}_1 d{\bf{x}}_2...d{\bf{x}}_{j-1}d{\bf{x}}_{j+1}...d{\bf{x}}_{N}$.
Now we can find the following coordinate-independent value:
\begin{eqnarray}
\label{10}
\langle S(\gb) \rangle \langle S^{*}(\gb) \rangle =
\sum_{j=1}^{N} \int_V e^{-i{\bf{x}}_j \gb }{\cal{F}}_j({\bf{x}}_j)d{\bf{x}}_j
\int_V e^{i{\bf{x}}_j \gb}{\cal{F}}_j({\bf{x}}_j)d{\bf{x}}_j
+ 
\sum_{i,j=1, i\ne j}^{N}
e^{i({\bf{r}}_i-{\bf{r}}_j) \gb}
\int_V
e^{i({\bf{x}}_j-{\bf{x}}_i) \gb}
{\cal{F}}_i({\bf{x}}_i){\cal{F}}_j({\bf{x}}_j)
d{\bf{x}}_i d{\bf{x}}_j
\end{eqnarray}

In a similar manner, one can define the average square module of the
structure factor
\begin{eqnarray}
\label{11}
\langle S(\gb) S^*(\gb) \rangle =
\int S(\gb,{\bf{r}}_1 - {\bf{x}}_1,..,
{\bf{r}}_{N}- {\bf{x}}_{N})
S^*(\gb,{\bf{r}}_1 - {\bf{x}}_1,..,
{\bf{r}}_{N}- {\bf{x}}_{N})
{\cal{P}}
({\bf{x}}_1,{\bf{x}}_2,..., {\bf{x}}_{N})
d{\cal{V}} = \nonumber \\
N +\sum_{i,j=1,i\ne j}^{N}
e^{i({\bf{r}}_i-{\bf{r}}_j)\gb}
\int_V
e^{-i({\bf{x}}_i-{\bf{x}}_j)\gb}
{\cal{F}}_{ij}({\bf{x}}_i,{\bf{x}}_j)
d{\bf{x}}_i d{\bf{x}}_j\, ,
\end{eqnarray}
where
\begin{equation}
\label{12}
{\cal{F}}_{ij}({\bf{x}}_i,{\bf{x}}_j) =\int {\cal{P}} d{\cal{V}}(-i,-j)\, .
\end{equation}
Here  the term $d{\cal{V}}(-i,-j)$ indicates that the integration takes place over the
whole space variables, except the ones belonging to the $i,j$-atoms. 
Taking into account  the relations obtained here,  we can write
for the dispersion
\begin{eqnarray}
\label{13}
\langle \langle S(\gb) S^*(\gb) \rangle \rangle =
N -
\sum_{j=1}^{N} \int_V e^{-i{\bf{x}}_j \gb}{\cal{F}}_j({\bf{x}}_j)d{\bf{x}}_j
\int_V e^{i{\bf{x}}_j\gb} {\cal{F}}_j({\bf{x}}_j)d{\bf{x}}_j
+ \nonumber \\
\sum_{i,j=1,i\ne j}^{N}
e^{i({\bf{r}}_i-{\bf{r}}_j)\gb}
\int_V
e^{-i({\bf{x}}_i-{\bf{x}}_j)\gb}
[{\cal{F}}_{ij}({\bf{x}}_i,{\bf{x}}_j)
-{\cal{F}}_i({\bf{x}}_i ) {\cal{F}}_j({\bf{x}}_j)]
d{\bf{x}}_i d{\bf{x}}_j,
\end{eqnarray}
where we introduced the following notation:
$\langle \langle S(\gb) S^*(\gb) \rangle \rangle =
\langle S(\gb) S^*(\gb) \rangle
-\langle S(\gb)\rangle \langle S^*(\gb) \rangle $.

In the case when the $x_i$ and $x_j$ variables are statistically
independent, the following relation takes place:
${\cal{F}}_{ij}({\bf{x}}_i,{\bf{x}}_j)=
{\cal{F}}_i({\bf{x}}_i){\cal{F}}_j({\bf{x}}_j)$.
If all $N$-atoms are  statistically independent, the normalized
probability density function may be represented as
${\cal{P}}=\prod_{j=1}^{N}{\cal{F}}_j({\bf{x}}_j)$
and Eq. (13)  can be rewritten in the following form:
\begin{eqnarray}
\label{14}
\langle \langle S(\gb) S^*(\gb) \rangle \rangle =
N- \sum_{j=1}^{N} \int_V e^{-i{\bf{x}}_j \gb}
{\cal{F}}_j({\bf{x}}_j) d{\bf{x}}_j
\int_V e^{i{\bf{x}}_j \gb}
{\cal{F}}_j({\bf{x}}_j) d{\bf{x}}_j
\end{eqnarray}
In the case when all the atoms in the fundamental cell are equivalent,
this equation has the following simple form:
\begin{equation}
\label{15}
\langle \langle S(\gb) S^*(\gb) \rangle \rangle =
N(1-\langle s(\gb) \rangle  \langle s^*(\gb)\rangle ),
\end{equation}
where $\langle s \rangle$ is the averaged function
\begin{equation}
\label{16}
\langle s(\gb) \rangle =
e^{{i{\bf{r}}}\gb} \int_V e^{{-i{\bf{x}}}\gb}
{\cal{F}}({\bf{x}}) d{\bf{x}}.
\end{equation}

It should be noted that, for an ideal atomic  structure,
${\cal{F}}_j({\bf{x}}_j) = \delta({\bf{x}}_j)$ for every $j$
and then
 $\langle \langle S(\gb) S^*(\gb) \rangle \rangle  =0$.

It is significant that the equations in this section
are valid for any vector $\gb$ of the reciprocal space, in particular for
any vector of the reciprocal lattice. Besides, there are no indications
of the atomic sorts in the equations. Below, it will be shown that the
cross section of the coherent bremsstrahlung  depends on the
averaged structure factors for monoatomic structures and on
 some combinations of  similar factors for multiatomic structures.
Thus,  Eqs. ({8})-({16}) are also useful in the case of   structures
consisting of different atoms.

\section{Simulation of fluctuations}
Let us build the three-dimensional (${\cal{N}} \times {\cal{N}} \times {\cal{N}}$)
cubic lattice consisting of identical cubic
cells with side size equal to $a$. Then we can put into every cell
an identical number of atoms, which we denote by $N$. Doing this, it
would mean  knowing the coordinates of every atoms in every
cell. Let us select a local Cartesian coordinate system in every cell.
Besides,  the probability function of localization of atoms in the cell
will be  considered as known
 ${\cal{P}} ({\bf{x}}_1,{\bf{x}}_2,...,{\bf{x}}_{N})$.
Let us suppose that the atomic coordinates in every cell are distributed
according to this function.

To be specific, we select the basic Cartesian coordinate system in
the left and bottom  corner of the lattice. With this lattice
(of ${\cal{N}}a \times {\cal{N}}a \times {\cal{N}}a$ size) as the basic element, we can
build (by using  parallel translations in $x, y, z$-directions
with a period of ${\cal{N}}a$) the three-dimensional infinite periodic
(in the strict sense) structure.
The main idea in our consideration stems from the fact that
the above described superlattice (at large enough {\cal{N}})
contains practically all
combinations of atoms in the small cells (according to
the normalized probability density function ${\cal{P}}$) and, on the
other hand, this superlattice is periodic in a strict sense.
Thus, we can use the coherent bremsstrahlung theory for
describing the radiation processes.
For this purpose,
 the structure factors  of the large lattice should be found for every
reciprocal vector $\gbt$. Further we will introduce a tilda symbol                              
above the  values relative to the large cube lattice with the side of $Na$.
Thus, these structure factors
$\tilde{S}$  can be calculated from the following relation:
\begin{equation}
\label{17}
  \tilde{S}(\gbt) = \sum_{i=1}^{\cal{N}} \sum_{j=1}^{\cal{N}} \sum_{k=1}^{\cal{N}}
 e^{{i \gbt {\bf{ r_{ijk}}}}} S_{ijk}(\gbt)
\end{equation}
where the reciprocal vector $\gbt$ is
\begin{equation}
\label{18}
\gbt = G_0 l {\bf{e}}_x  + G_0 m {\bf{e}}_y +G_0 n {\bf{e}}_z
 , \, \,\, l,m,n = 0,\pm 1, \pm 2, ...
\end{equation}
Here  $G_0= 2\pi/({\cal{N}}a)$, ${\bf{e}}_x$, ${\bf{e}}_y$ and
${\bf{e}}_z$ are the unit vectors in $x,y,z$-directions
 and the translation vector reads ${\bf{r}}_{ijk} = (i-1) a {\bf{e}}_x  +
(j-1) a{\bf{ e}}_y +(k-1) a{\bf{e}}_z $
in the basic coordinate system.
In Eq. (17)  $S_{ijk}(\tilde{g})$ denotes the  following structure
factor:  $\sum_{\alpha=1}^{N} \exp{i \gbt{{\bf{r}}}_{ijk\alpha }}$,
where ${{\bf{r}}}_{ijk\alpha}$ is the vector-radius of
the $\alpha$-th atom
in the local coordinate system of $ijk$-cell.
The multiplication $\gbt{\bf{ r_{ijk}}}$ reads
\begin{equation}
\label{19}
\gbt{\bf{ r_{ijk}}} = 2\pi( {l(i-1) \over {\cal{N}}} + {m(j-1) \over {\cal{ N}}} + {n(k-1) \over {\cal{ N}}}).
\end{equation}
For large enough numbers ${\cal{N}}$ one can find
\begin{equation}
\label{20}
\tilde S(\gbt) =\sum_{i=1}^{\cal{N}} \sum_{j=1}^{\cal{N}} \sum_{k=1}^{\cal{N}}
 S_{ijk}(\gbt) \approx {\cal{N}}^3 \langle S(\gbt) \rangle,
\end{equation}
where $\langle S(\gbt) \rangle$ is the structure factor $S_{ijk}$ averaged over the coordinates
and the quantities $l/{\cal{N}},m/{\cal{N}}, n/{\cal{N}}$ are  integer numbers.
In the other cases, one can write
\begin{equation}
\label{21}
 \tilde{S}(\gbt) \rightarrow 0 \qquad at \qquad {\cal{N}} \rightarrow \infty \,.
\end{equation}
Eq. (20) is obvious, so at the pointed condition the value $\gbt{\bf{ r_{ijk}}}$
is multiplied by $2\pi$ and  the exponents in Eq. (20) are equal to 1.
For obtaining Eq. (21) we should take into account that
there is only a finite number of the various exponents in Eq. (17) (this
number $N_{exp} \le {\cal{N}}$, the equality holding when ${\cal{N}}$ is a prime number).
 Grouping the terms near the same exponents, we get
\begin{equation}
\label{22}
\tilde{S}(\gbt) = \sum_{p=1}^{N_{exp}} (e^{2\pi p /N_{exp}} \sum {S_{i_p}}) \approx
\sum {S_{i_p}}  \sum_{p=1}^{N_{exp}} e^{2\pi p /N_{exp}} = 0 \,.
\end{equation}
Here we can remove $\sum {S_{i_p}}$-terms, due to their approximated
equality at large enough ${\cal{N}}$.

We can consider the structure factor $\tilde S(\gbt)$
as a statistical variable. Taking into account Eqs. (20),(21), we
find that $\langle \tilde S(\gbt) \rangle =
{\cal{N}}^3 \langle S(\gbt) \rangle$ when $l/{\cal{N}}, m/{\cal{N}}, n/{\cal{N}}$ are
simultaneously integer numbers and $\langle \tilde S(\gbt) \rangle = 0$
in the other cases.

Similar calculations allow us to obtain the averaged dispersion 
of the structure factor $\tilde{S}(\gbt)$ for the case
of large ${\cal{N}}$-numbers
\begin{equation}
\label{23}
\langle \langle \tilde{S}(\gbt) \tilde{S}^*(\gbt) \rangle \rangle=
{\cal{N}}^3 \langle \langle S(\gbt) S^*(\gbt) \rangle \rangle \,.
\end{equation}
This equation is valid for arbitrary $l,m,n$-numbers. However,
the value $\tilde{S}$ is defined by the limit of Eq. (20) in the case
when $l/{\cal{N}},m/{\cal{N}},n/{\cal{N}}$ are integer numbers, and $\langle \tilde S\rangle =0$ in
the other cases (see Eq. (21)).

It should be noted that our previous consideration is based on
the specific kind of the crystallographic structure.
It is easy to verify that our analysis is valid in the general
case. Indeed, the important relation for the correctness of the theory
(see Eq. ({19})) is valid for any real crystallographic
structure \cite{Kit, LL}.
Of course, in the general case, the
vector of the reciprocal lattice and the translation vector
must be written in the corresponding (generally nonorthogonal)
coordinate system.

\section{Cross section}
In principle, now we can calculate the
coherent bremsstrahlung cross section for the structure described
in the previous section.
However, we call to the attention of the reader the fact that the structure factors $\tilde{S}$,
in the case when the $l/{\cal{N}},m/{\cal{N}}, n/{\cal{N}}$-numbers are nonintegral ones,
tend to zero at large ${\cal{N}}$. Nevertheless, the contribution from
these factors to the calculated values may be noticeable, due to
their large amount.

In this case, the differential cross section of coherent bremsstrahlung
has the following form:
\begin{equation}
\label{24}
d\sigma_{CB}={(2\pi)^3 \over {{\cal{N}}^3N \tilde{V}}}
\sum_{\gbt} d\sigma_{BG}
\tilde{S}(\gbt)
\tilde{S^*}(\gbt)
\delta({\bf{q}}/\hbar -\gbt) \,.
\end{equation}
Here $\tilde{V}= {\cal{N}}^3V$ is the volume of the fundamental cell of the structure.

When the number ${\cal{N}}$ is large enough, we can write (see Eqs. ({20})-({23}))
\begin{eqnarray}
\label{25}
\tilde S(\gbt)\tilde S^*(\gbt) \approx
\langle \tilde S(\gbt)\tilde{S}^*(\gbt)\rangle=
\langle \tilde S(\gbt)\rangle \langle\tilde S^*(\gbt)\rangle
+{\cal{N}}^3\langle \langle  S(\gbt){S}^*(\gbt)\rangle \rangle
\end{eqnarray}
for the case of the simultaneously integer numbers $l/{\cal{N}},m/{\cal{N}}, n/{\cal{N}}$
and
\begin{eqnarray}
\label{26}
\tilde S(\gbt)\tilde S^*(\gbt) \approx
 \langle \tilde S(\gbt)\tilde S^*(\gbt)\rangle=
{\cal{N}}^3\langle \langle  S(\gbt){S}^*(\gbt)\rangle \rangle
\end{eqnarray}
for the other cases.
Taking into account these relations and Eq. (20), we get
\begin{eqnarray}
\label{27}
d\sigma_{CB} = {(2\pi)^3 \over {N V}}
\sum_{\gb}d\sigma_{BG}
\langle S(\gb)\rangle \langle S^{*}(\gb)\rangle
\delta({\bf{q}}/\hbar -\gb)
+ 
{(2\pi)^3 \over {N \tilde{V}}}
\sum_{\gbt}d\sigma_{BG}
\langle \langle  S(\gbt){S}^*(\gbt)\rangle \rangle
\delta({\bf{q}}/\hbar -\gbt)\,.
\end{eqnarray}
We see that the total cross section represents the sum of
two  terms.  We interpret the first term as the averaged   coherent
contribution. Indeed, the value of this cross section calculated
per atom is proportional to N  and inversely proportional to V.
In addition, this cross section is independent of ${\cal{N}}$, which
defines the size of the large lattice.

It is obvious that the second term describes the common effect due to
 the incoherent bremsstrahlung
in the ${\cal{N}} \times {\cal{N}} \times {\cal{N}}$ lattice at large ${\cal{N}}$-values
and the coherent one in the infinite superlattice.
Now we can find the energetic cross section by integration over ${\bf{q}}$
\begin{eqnarray}
\label{28}
d\sigma_{CB} = d E_\gamma \lbrace \int_{{\bf{q}}_{m}}^{\infty}
{(2\pi)^3 \over {N V}}
\sum_{\gb}d\sigma_{BG}
\langle S(\gb)\rangle \langle S^{*}(\gb)\rangle
\delta({\bf{q}}/\hbar -\gb)  d{\bf{q}}
+
{{(2\pi\hbar)}^3\over N{\tilde{V}}} \sum_{\gbt}
d\sigma_{BG}(E, E_\gamma, \hbar\gbt)
\langle \langle S(\gbt) S^*({\gbt) \rangle \rangle
}\rbrace\, ,
\end{eqnarray}
where ${\bf{q}}_m$ is the vector  directed along the velocity of the electron beam and
$ q_m= \hbar\delta+q_{\bot}^2c/E$, $q_\bot$ is the transversal projection
of ${\bf{q}}$ on the primary electron direction of motion, $m$  is the electron mass,
$c$ is the speed of light, and for
the minimal value of the transfer momentum $\hbar\delta$ we use the traditional
notation.

We see that the second term in Eq. ({28}) is proportional to a sum over reciprocal
vectors $\gbt$ (see Eq. ({18})).
At large {\cal{N}}, this sum is most conveniently expressed by an integral representation.
For this purpose, we use the relation: $dldmdn= d\gbt/G_0^3 = d{\bf{q}} /(\hbar G_0^3)$.
As a result, we get
\begin{eqnarray}
\label{29}
d\sigma_{CB} = d E_\gamma \lbrace \int_{{\bf{q}}_{m}}^{\infty}
{(2\pi)^3 \over {N V}}
\sum_{\gb}d\sigma_{BG}
\langle S(\gb)\rangle \langle S^{*}(\gb)\rangle
\delta({\bf{q}}/\hbar -\gb)  d{\bf{q}}
+
{1\over N} \int_{{\bf{q}}_{m}}^{\infty}
d\sigma_{BG}
\langle \langle S({\bf{q}}) S^*({\bf{q}}) \rangle \rangle
d{\bf{q}}
\rbrace \,.
\end{eqnarray}

We stress that the structure factors in the first term 
(see Eq. ({29 })
 are discrete values,
which depend on
the reciprocal lattice vectors $\gb$, and that
 the structure factor   $S({\bf{q}})$ in the second term is a continuous
function of the $q/\hbar$ variable.
Note that passing from a discrete to a
continuous description, one removes the action of the large periodic
lattice.

Thus, in principle, we have solved the problem of the coherent bremsstrahlung in 
imperfect periodic atomic  structures. In fact,
Eq. ({29}) represents the sum of the coherent and incoherent contributions
in the cross section: $ d\sigma_{BC}= d\sigma_{c} + d\sigma_{i}$.
Now  both cross sections should be reduced to a form, which is convenient
for specific calculations.

First of all, we find the coherent contribution.
For this purpose, it is necessary to simplify the cross section $d\sigma_{BG}$
as it was described in Ref.  \cite{TM}. This simplification is based on the
fact that the effective range of $q \ll mc $, due to thermal fluctuations.

Using this condition, and with the help of  calculations similar to those in Ref. \cite{TM,Tim},
 we can obtain for the coherent cross section
\begin{equation}
\label{30}
x{d\sigma_c \over dx} =
\sigma_0 [(1+(1-x)^2)\psi_1 - {2\over 3}(1-x)\psi_2],
\end{equation}
where $\sigma_0= \alpha_{QED} Z^2 r_e^2$, $\alpha_{QED} =1/137.04$,
$r_e$ is the classical electron radius,
 $x={ E_{\gamma}\over E}$ is
the ratio of the emitted photon energy $E_{\gamma}$ to the initial energy $E$ of the electron,
and $\psi_1,\,\psi_2$-functions are
\begin{equation}
\label{31}
\psi_1=4 {(2\pi)^2 \over NV } \sum_{\bf g}|{U({\bf g})}|^2
{\delta \gm^2_{\bot} \over \gm_{||}^2},
\end{equation}
\begin{equation}
\label{32}
\psi_2= 24{(2\pi)^2 \over NV }\sum_{\bf g}|{U({\bf g})}|^2
{\delta^2 \gm^2_{\bot}(\gm_{||}-\delta) \over \gm_{||}^4}.
\end{equation}
Here
 $\gb$ is the vector of the reciprocal lattice,
 $\gm_{||}$ is the projection of the $\bf{g}$-vector on the direction of the
particle motion, $\gm^2_{\bot}=\gm^2-\gm_{||}^2$, and the $\hbar\delta$-value is given by
\begin{equation}
\label{33}
\hbar \delta = {m^2c^3 \over 2E } {x \over 1-x} \,.
\end{equation}
The summation in Eqs. ({31}),({32}) is carried out under the following condition:
\begin{equation}
\label{34}
 \gm_{||} \ge \delta \,.
\end{equation}
The $|U(\gb)|^2$- values are
\begin{equation}
\label{35}
|U({\bf g})|^2=\langle S({\bf g})\rangle \langle S^*({\bf g})\rangle {\frac{{(1-F(\gm))^2}}{{\gm^4}}},
\end{equation}
where $F(\gm)$ is the atomic form factor. These equations differ from the standard theory
by the averaged structure factor in Eq. ({35}).

Next, we simplify the second integral in Eq. ({29}) describing
the incoherent contribution.
From Eq. ({14}) we see that the incoherent cross section
may be represented as the difference $(d\sigma_{BG} - d\sigma_{d})$,
where $d\sigma_{BG}$ is the cross section for the process of bremsstrahlung
in the corresponding amorphous media and $d\sigma_{d}$ is the cross
section depending on the averaged structure factors. This allows us to
simplify Eq. ({29})
\begin{equation}
\label{36}
x{d\sigma_i \over dx} =
\sigma_0 [(1+(1-x)^2)\psi^a_{1s} - {2\over 3}(1-x)\psi^a_{2s}],
\end{equation}
where the functions $\psi^a_{1s},\psi^a_{2s}$ have the following form:
\begin{equation}
\label{37}
\psi^a_{1s}=\psi^a_{1BG}- \psi^a_{1d},\,\psi^a_{2s}=\psi^a_{2BG}- \psi^a_{1d}\,.
\end{equation}
\begin{equation}
\label{38}
\psi^a_{1d} = {2 \hbar \delta \over N} \int_0^{\infty}dq^2_{\bot}
\int_{\hbar\delta}^{\sim mc} [\langle \langle S({\bf{q}}) S^*({\bf{q}})
 \rangle \rangle - N ]
{[1-F(q)]^2 \over q^4} {q_{\bot}^2 \over q^2_{\|}}dq_{\|},
\end{equation}
\begin{eqnarray}
\label{39}
\psi^a_{2d} = {12 (\hbar \delta)^2 \over N} \int_0^{\infty}dq^2_{\bot}
\int_{\hbar \delta}^{\sim mc} [\langle \langle S({\bf{q}}) S^*({\bf{q}}) \rangle \rangle - N ]
{[1-F(q)]^2 \over q^4} {q_{\bot}^2(q_{\|}-\hbar\delta)
 \over q^4_{\|}}dq_{\|},
\end{eqnarray}
and the well-known \cite{TM, Tim} functions $\psi^a_{1BG}$ and $\psi^a_{2BG}$
are
\begin{equation}
\label{40}
\psi^a_{1BG}(\delta)= 4+4\int_{\hbar\delta}^{\sim mc}
(q-\hbar\delta)^2{[1-F(q)]^2 \over q^3}dq,
\end{equation}
\begin{equation}
\label{41}
\psi^a_{2BG}(\delta)= 10/3 +4\int_{\hbar\delta}^{\sim mc}
(q^3-6\hbar^2\delta^2q \ln({q \over \hbar \delta })
+3\hbar^2\delta^2q -4\hbar^3 \delta^3
){[1-F(q)]^2 \over q^4}dq.
\end{equation}
 Eqs. ({38}),({39}) were obtained under the assumption $q \ll mc$, which is always
valid for real atomic structures.
In the case when the $S$-factors are functions only of the
 $q$-variable, Eqs. ({38}),({39}) are simplified
\begin{equation}
\label{42}
\psi^a_{1d} =
4\int_{\hbar\delta}^{\sim mc}
(q-\hbar\delta)^2[\langle \langle S({q}) S^*({q}) \rangle \rangle/N -1]
{[1-F(q)]^2 \over q^3}dq,
\end{equation}
\begin{equation}
\label{43}
\psi^a_{2d} =
4\int_{\hbar\delta}^{\sim mc}
(q^3-6\hbar^2\delta^2q \ln({q \over \hbar \delta })
+3\hbar^2\delta^2q -4\hbar^3 \delta^3)
[\langle \langle S({q}) S^*({q}) \rangle \rangle /N -1]
{[1-F(q)]^2 \over q^4}dq\,.
\end{equation}
When the condition of complete screening is fulfilled, Eqs. ({40}),({41})
take the following simple form:
\begin{equation}
\label{44}
\psi^a_{1BG}= 4\ln 183 Z^{-{1 \over 3}}\,, \quad
\psi^a_{2BG}= 4\ln 183 Z^{-{1 \over 3}} -2/3\,.
\end{equation}

Note that the cross section $d\sigma_{BC}$  can be found in
another way
with the help of the Fourier transform (see Ref. \cite{TM}). However,
this method is longer and requires straightforward but cumbersome
calculations.
\section{Averaging over thermal fluctuations}
The relations obtained in this paper  describe the process of
coherent bremsstrahlung in  imperfect periodic structures.
One can see that, in the general case, the cross section is the
sum  of coherent and incoherent contributions.
Thermal atomic fluctuations always take place in
atomic structures. Let us apply our theory for the study of their influence
on the coherent bremsstrahlung spectrum.
Here we take into account the simplest  case, when the thermal
fluctuations are isotropic in  space and independent of
the location of other  atoms. Then, in accordance with Ref. \cite{TM, BKS}
\begin{equation}
\label{45}
{\cal{F}}_j({\bf{x}}_j)= {\exp(-x_j^2/(2A))\over (2\pi A)^{3/2}} \,,
\end{equation}
and ${\cal{F}}_{ij}({\bf{x}}_i,{\bf{x}}_j)=
{\cal{F}}_i({\bf{x}}_i){\cal{F}}_j({\bf{x}}_j)$.
The normalized probability density function is equal to the product of all
${\cal{F}}_j$-functions.
From Eqs. ({8})-({16}) we find
\begin{equation}
\label{46}
\langle S(\gb)\rangle=\sum_{j=1}^{N}
 e^{ i {\bf{r}}_{j} \gm }
 e^{-A \gm^2/2} = S(\gb)e^{-A \gm^2/2} \,,
\end{equation}
\begin{equation}
\label{47}
\langle S(\gb)\rangle \langle S^*(\gb) \rangle =S(\gb)S^*(\gb)e^{-A \gm^2} \,,
\end{equation}
\begin{equation}
\label{48}
\langle \langle S(\gb) S^{*}(\gb) \rangle \rangle
 =
N (1 -e^{-A\gm^2}) \,.
\end{equation}
In order to take into account these fluctuations,
we need to substitute $\langle S \rangle$ and
$\langle SS^{*}\rangle$-values
in Eqs. ({35}),({42}),({43}). In this case the relations for the incoherent part
of the cross section are in the agreement with similar
ones in Ref. \cite{TM,U,Tim,BKS}.

One can assume that in  most cases the mechanism violating
the ideal structure acts independently of the thermal fluctuations.
Then the normalized probability density function ${\cal{P}}$ of the structure
may be written in the following form:
${\cal{P}}({\bf{r}}_1,...,{\bf{r}}_{N},
{\bf{r}}_1^T,...,{\bf{r}}_{N}^T) =
{\cal{P}}_C({\bf{r}}_1,...,{\bf{r}}_{N})
{\cal{P}}_T({\bf{r}}_1^T,...,{\bf{r}}_{N}^T)
$
where ${\cal{P}}_T$ is the normalized probability density function for thermal fluctuations
(see Eq. ({46})), and ${\cal{P}}_C$ is some other similar function.
Let us consider the case when the atomic system is described by Eq. ({46}),
or, in other words, we assume that all atoms in the
fundamental cell are equivalent, with respect to thermal fluctuations.
Then, we get
\begin{equation}
\label{49}
\langle S \rangle_{TC} = e^{-A\gm^2/2} \langle S \rangle_C \,,
\end{equation}
\begin{eqnarray}
\label{50}
\langle \langle SS^* \rangle \rangle_{TC}=
N-(N-\langle \langle SS^*\rangle \rangle_C)e^{-A\gm^2} \,,
\end{eqnarray}
where the symbols T and C denote the corresponding averaging.
Under a similar assumption, that all atoms in the cell
are equivalent, we get instead of Eq. ({50})
\begin{equation}
\label{51}
\langle \langle SS^* \rangle \rangle_{TC}=
 N(1 - e^{-A\gm^2}\langle s \rangle \langle s^* \rangle) \,.
\end{equation}

The equations obtained here  may be substituted into Eqs. ({35}),({42}),({43}) and
 the problem of calculating the coherent bremsstrahlung for atomic systems, with the conditions
 pointed out above, is solved.

\section{Coherent bremsstrahlung in diatomic single crystals}
It is well-known that the process of  coherent bremsstrahlung may be
considered as a result of the electron motion in a
continuous periodic potential\cite{BKS}. In the case of  imperfect periodic
structures, we can also write the effective averaged potential
\begin{equation}
\label{52}
\varphi ({\bf r})={\frac{{4\pi eZ}} V }\sum_{{\bf g}}\langle S({\bf g}) \rangle
{(1-F(\gm))\over \gm^2} e^{-i{\bf
gr}} \,,
\end{equation}
where $\langle S({\bf g}) \rangle $ is the averaged
structure factor. From Eq. ({52}) one can also get one and two-dimensional
potentials (see \cite{BKS,MV1}).

As it can be seen from Eq. ({52}), our consideration is valid for  monoatomic
single crystals. However, the case of the polyatomic periodic structure
may be studied in the similar manner. Let us consider the diatomic
perfect periodic structure.
We can represent this structure as  the sum of two independent structures,
one of them consisting  of atoms with $Z_1$-number, and the other consisting
of atoms with $Z_2$-number. Both structures have the same periods and the
crystallographic  type of the three-dimensional lattice. We can write the three-dimensional
potential for this structure as
\begin{equation}
\label{53}
\varphi ({\bf r})={\frac{{4\pi e}} V }\sum_{{\bf g}}
{[Z_1 S(Z_1,{\bf g})(1-F(Z_1,\gm)) + Z_2 S(Z_2,{\bf{g}})(1-F(Z_2,\gm))]
\over \gm^2} e^{-i{\bf
gr}} \,,
\end{equation}
where $F(Z_1,\gm),\, F(Z_2,\gm)$ are the corresponding atomic form factors
and $S(Z_1,{\bf g})$,  $S(Z_2,{\bf g})$ are the structure factors
for every sublattice. They have  a form as in Eq. ({6}):
$S(Z_1,\gb)= \sum_{j=1}^{N_1} \exp{ir_j\gb},\,\,
S(Z_2,\gb)= \sum_{j=1}^{N_2} \exp{ir_j\gb}$, but the sum
should be taken  separately over
atoms of each sort, these numbers being denoted as $N_1,\, N_2$. The total number
of atoms in the fundamental cell is equal to $N=N_1+N_2$.

Then, we should  take into account that the bremsstrahlung
scattering amplitude  is proportional to the
Fourier transform of the
potential (see Eq. ({53})) and the cross section is proportional to the squared
amplitude. Thus, we can get the cross section of the process in the perfect
periodic structure, which is
defined by the following factor:
\begin{equation}
\label{54}
Y(Z_1,Z_2,{\bf g})= [Z_1 S(Z_1,{\bf g})(1-F(Z_1,\gm))
+ Z_2 S(Z_2,{\bf{g}})(1-F(Z_2,\gm))].
\end{equation}
The coherent bremsstrahlung cross section is proportional to the $YY^*$-value.

With the help of the above-considered method one can get the corresponding
cross section for the imperfect periodic diatomic structures. In the general
case, the function ${\cal{P}}$ contains the space variables for every
atom in the fundamental cell, and different correlations between
various atoms are possible, in principle. Below, we will write  the final
result
for coherent bremsstrahlung in the
 diatomic structures, taking into account
thermal fluctuations. We carry out our calculations, under the assumption that
fluctuations of all atoms are isotropic and independent,
but the squared radius of the vibrations depends on the sort of atoms.
The final result for the cross section, calculated per fundamental cell,
has the following form:
\begin{equation}
\label{55}
x {d\sigma \over dx} = \alpha_{QED} r_e^2
[(1+(1-x)^2)(\psi_1 + \psi_1^a) -{2 \over 3}(1-x)(\psi_2 + \psi_2^a)] \,,
\end{equation}
where
\begin{equation}
\label{56}
 \psi_1= 4{(2\pi)^2 \over V}
\sum_{\gb}
\langle Y(\gb)\rangle \langle Y^*(\gb) \rangle
{\delta \gm_\bot^2 \over \gm^4 \gm^2_{||}} \,,
\end{equation}
\begin{equation}
\label{57}
 \psi_2= 24{(2\pi)^2 \over V}
\sum_{{\bf{g}}}
\langle Y(\gb)\rangle \langle Y^*(\gb) \rangle {\delta^2 \gm_\bot^2(\gm_{||}- \delta) \over \gm^4 \gm^4_{||}} \,.
\end{equation}
The summation in Eqs. ({56}),({57}) is carried out with the condition $\gm_{||} \ge \delta$.
The functions $\psi_1^a, \psi_2^a$ are calculated according to
\begin{equation}
\label{58}
\psi_1^a = N_{1}Z^2_1 \psi_{1BG}(Z_1)^a + N_{2}Z^2_2 \psi_{1BG}(Z_2) -\psi_{1d}^a \,,
\end{equation}
\begin{equation}
\label{59}
\psi_2^a = N_{1}Z^2_1 \psi_{2BG}(Z_1)^a + N_{2}Z_2^2 \psi_{2BG}(Z_2) -\psi_{2d}^a \,,
\end{equation}
where
\begin{equation}
\label{60}
\psi^a_{1d} =
4\int_{\hbar\delta}^{\sim mc}
(q-\hbar\delta)^2[\langle \langle Y(q)Y^*(q)\rangle \rangle
 -N_{1}Z^2_1(1-F(Z_1))^2-N_{2}Z^2_2(1-F(Z_2))^2]
{dq \over q^3},
\end{equation}
\begin{equation}
\label{61}
\psi^a_{2d} =
4\int_{\hbar\delta}^{\sim mc}
(q^3-6\hbar^2\delta^2q \ln({q \over \hbar \delta })
+3\hbar^2\delta^2q -4\hbar^3 \delta^3)
[\langle \langle Y(q)Y^*(q)\rangle \rangle -
N_{1}Z^2_1(1-F(Z_1))^2-N_{2}Z^2_2(1-F(Z_2))^2]
{dq \over q^4},
\end{equation}
with $\langle \langle Y(q)Y^*(q)\rangle \rangle =
\langle Y(q)Y^*(q)\rangle -\langle Y(q)\rangle \langle Y^*(q)\rangle$.

In the case of thermal fluctuations, one can find
\begin{equation}
\label{62}
\langle Y(\gb)  \rangle =
Z_1  S(Z_1.\gb) (1-F(Z_1,\gb))e^{-A_1 \gm^2/2} + Z_2  S(Z_2,\gb) (1-F(Z_2,\gb))e^{-A_2 \gm^2/2} \,,
\end{equation}
\begin{eqnarray}
\label{63}
\langle \langle Y(\gb)Y^*(\gb)\rangle \rangle =
N_{1}Z_1^2 (1-F(Z_1,\gb))^2(1-e^{-A_1 \gm^2}) +
N_{2}Z_2^2 (1-F(Z_2,\gb))^2(1-e^{-A_2 \gm^2}) \,,
\end{eqnarray}
where $A_1$ and $A_2$ are the squared radii of the thermal vibrations,
for the first and second sorts of atoms, respectively.
We recall that, in Eqs. (62),(63),  the variable $\gb$ is discrete in the calculation of the
coherent contribution and continuous for the incoherent one.

In a similar manner one can calculate the cross section
for periodic structures consisting of three and more atoms.
Note that the total intensity radiation per unit of length in multiatomic structures may be
calculated as $ N_c E \int_0^1 x {d\sigma \over dx}dx$, where
${d\sigma \over dx}$
is defined by Eq. ({55}) and $N_c= 1/V$ is the number of fundamental cells per
unit  volume.

\section{Examples of calculations}
\subsection{Limiting cases of atomic structures}
Let us consider an ideal atomic structure. Obviously, the  ${\cal{P}}$-function
for this structure is given by the following  multiplication:
\begin{equation}
\label{64}
{\cal{P}}({\bf{x}}_1,{\bf{x}}_2,...,{\bf{x}}_{N})=
\prod_{j=1}^{N} \delta({{\bf{x}}_j}) \,.
\end{equation}
Taking this fact into account, we get
 $\langle S({\bf{g}}) S^*({\bf{g}}) \rangle
-\langle S({\bf{g}})\rangle \langle S^*({\bf{g}}) \rangle  =0$.
This means that, for an ideal atomic structure, the incoherent contribution
in the cross section is equal to zero.

Now we consider a monoatomic homogeneous amorphous medium.
We can find the mean volume $V = a^3$ per atom and build the
cubic fundamental cell, which contains $N$ atoms. This means that
the cube side of this cell is equal to $N^{1/3}a$. We take
the ${\cal{P}}$-function as a product of the following functions
defined on the whole volume of the fundamental cell:
\begin{equation}
\label{65}
{\cal{P}}= \prod_{j=1}^{N}{\cal{F}}_j({\bf{x}_j}), \qquad
{\cal{F}}_j({\bf{x}}_j)={1\over 8b^3},  \qquad -b \le x_{ji} \le b, \, i=1,2,3,\,
\,\, b= N^{1/3}a/2.
\end{equation}
The averaged structure factor (see Eq. (16)) for an atom in the cell is
\begin{equation}
\label{66}
\langle s( {\bf{q}}) \rangle
= 8{ \sin(N^{1/3}a q_1/2) \sin(N^{1/3}a q_2/2)\sin(N^{1/3}a q_3/2)
\over N a^3 q_1q_2q_3}\,.
\end{equation}
In general, we consider the variable ${\bf{q}}$ in the latter equation
as a continuous one. However, for the calculation of the coherent contribution,
we should  take a discrete set of quantities of the variable, which
is described by a relation similar to Eq. ({18}). Substituting in Eq. ({66})
$q_j= {4\pi \over b} l_j$, $(l_j=1,2,3...)$ we get $\langle s( {\bf{q}}) \rangle =0$,
and according to Eq. (8)  every discrete structure factor S is also equal to 0.
This means that there is no  coherent contribution in the cross section.

Taking into account the calculations of the incoherent contribution
we should  consider the  $\langle s( {\bf{q}}) \rangle$-value as
a function of the continuous variable  ${\bf{q}}$. It is easy to see
that the $\langle s( {\bf{q}}) \rangle \langle s^*( {\bf{q}}) \rangle$-value
at small ${\bf{q}}$ is approximately equal to 1 and at large enough ${\bf{q}}$
this value is significantly less than 1. The larger are the numbers $N$, and
then the smaller are the
$q$-values at which this rule holds. Taking into account that there exists a minimum transfer momentum
in the bremsstrahlung process, we can select the value of $N$ such that
$\langle s( {\bf{q}}) \rangle \langle s^*( {\bf{q}}) \rangle \approx 0$,
and therefore
$\langle \langle S({\bf{q}})S^*({\bf{q}})\rangle \rangle
\approx N$.
In this case the incoherent contribution is the same as in the corresponding amorphous medium.
With the help of a similar function (see Eq. ({65})) one can describe  the transition
from  a three-dimensional structure to two-dimensional or one-dimensional ones.

\subsection{Nanotube superlattice}
Let us calculate the bremsstrahlung cross section
in the nanotube superlattice (see Fig. 1 and Eqs. ({1})-({3})), under
the assumption of a random distribution of the angle shifts.
In this case we rewrite Eqs. ({1})-({3})  in  cylindrical
coordinates $ \rho,\varphi, z$
\begin{eqnarray}
\label{67}
\rho_{1,j} = R \,\,\,\rho_{2,j} = R \,, \\
\label{68}
\varphi_{1,j}= \varphi_{1,1} +4\pi(j-1)/N,\,\,
\varphi_{2,j}= \varphi_{2,1} +4\pi(j-1)/N \,, \\
\label{69}
 z_{1,j}=0,\,\,\,\,  z_{2,j}=b/2 \,,
\end{eqnarray}
where $j= 1,...,N/2$ and $\varphi_{1,1} -\varphi_{2,1} =const$.
These equations describe the various nanotubes.
In particular, for a (10,10) armchair single nanotube, we have
$N=40,\varphi_{1,1} -\varphi_{2,1} = 4\pi/(3N)$
and the other geometric parameters are shown in Fig. 1.

It is easy to see that the problem of averaging has only one independent coordinate,
$\varphi_{1,1}$, for instance. In principle, for its solution one can get the  necessary
averaging values with the simple density function $1 \over 2\pi$. However, for the sake
of illustration, we begin with giving the ${\cal{P}}$-function
\begin{eqnarray}
\label{70}
{\cal{P}}= {1\over 2\pi}\delta(z_{1,1}){\delta(\rho_{1,1}-R)\over R}
 \prod_{j=2}^{N/2}\delta (\varphi_{1,j}-\varphi_{1,1}-{4\pi \over N}(j-1))
{\delta(z_{1,j})\delta(\rho_{1,j}-R)\over R} \nonumber \\
 \prod_{j=1}^{N/2}\delta (\varphi_{2,j}-\varphi_{2,1}-{4\pi \over N}(j-1))
{\delta(z_{2,j}-b/2)\delta(\rho_{2,j}-R)\over R}
\end{eqnarray}
with the unit of volume
\begin{equation}
\label{71}
d{\cal{V}}=\prod_{j=1}^{N/2}
\rho_{1,j} d\rho_{1,j} d\varphi_{1,j} dz_{1,j}
\prod_{j=1}^{N/2}
\rho_{2,j} d\rho_{2,j} d\varphi_{2,j} dz_{2,j} \,.
\end{equation}
From this we find the ${\cal{F}}_{ij}$-functions
\begin{equation}
\label{72}
{\cal{F}}_{[m,i],[n,j]}= {1\over 2\pi}
\delta(\varphi_{n,j}-\varphi_{m,i} + {4\pi \over N}(i-j) +\Delta_{n,m})
{\delta(z_{n,j}-b_{n,m})\delta(\rho_{n,j}-R)\over R}
{\delta(z_{m,i}-b_{n,m})\delta(\rho_{m,i}-R)\over R} \,.
\end{equation}
Here every atom is labelled by a pair of numbers $m,i$, where
$m= 1$ or $2$ is the number of the ring (see Fig. 1) and $i= 1,2,..N/2$
is the atomic number in the selected ring. The value $\Delta_{n,m}$ is equal to
$\varphi_{1,1} -\varphi_{2,1} = \Delta_{2,1}=-\Delta_{1,2}$
when $n \ne m$, and it vanishes when $n=m$, and the value $b_{m,n}$ is equal to 0 or $b$/2, in
accordance with Eq. (70).

From here, we can obtain by integration
\begin{equation}
\label{73}
{\cal{F}}_{[m,i]}={1\over 2\pi}{\delta(z_{m,i}-b_{n,m})\delta(\rho_{m,i}-R)\over R} \,,
\end{equation}

Now we can calculate the averaged structure factors
\begin{equation}
\label{74}
\langle  S({\gm_\bot},\gm_z) \rangle =
{N \over 2} J_0(R \gm_\bot)(1+e^{ib \gm_z /2}) \,.
\end{equation}
Here $J_0(x)$ is the Bessel function of the zero-th order and
$\gm_\bot,\gm_z $ are the
values of the reciprocal vector projection on the xy-plane and its projection
on the $z$-axis, respectively (see Fig. 1). Eq. ({74}) describes the continuous structure
factors. For the determination of the discrete set of the structure factors, needed for calculating
the coherent contribution (see Eqs. ({30})-({35})), it is necessary  to substitute
in Eq. ({74}) the projections of the reciprocal vectors
$\gb_\bot = {2\pi\over a} ((l-m){\bf{e}}_x +(l+m){\bf{e}}_y/ \sqrt{3})$,
$\gm_z ={2\pi \over b}n {\bf{e}}_z$ for the triangular crystallographic lattice.
As a result, we get $S(l,m,n) = N J_0(R\gm_\bot)$ for even n-numbers and
0 for odd ones.

Then we find
\begin{eqnarray}
\label{75}
\langle \langle {S}(\gb){S^{*}}(\gb)\rangle \rangle
=
2\sum_{\nu=1}^{N/2}\sum_{\eta=1}^{N/2}
[J_0(2\gm_{\bot}R\sin({4\pi \over N}(\nu-\eta)))\, +  \nonumber \\
J_0(2\gm_{\bot}R\sin({4\pi \over N}(\nu-\eta) +(\varphi_2- \varphi_1)))
 \cos(\gm_zb/2)]
- N^2J_0^2(\gm_\bot R)(1+\cos(\gm_zb/2))/2 \,.
\end{eqnarray}

From these equations one can see that the obtained averaged
structure factors are  functions of the $q_\bot$ and $q_z$ variables.
This means that it is necessary to use Eqs. ({38}),({39}) in the calculations.
We compute first the internal integral, which is given by
\begin{equation}
\label{76}
I(q, \hbar \delta)= \int_{0}^{q^2-(\hbar\delta)^2}
{(\langle \langle{S}({\bf{q}}){S^{*}}({\bf{q}})\rangle \rangle
/N -1)
q^2_{\bot}dq_\bot^2
\over
(q^2-q^2_\bot)^{3\over 2}}\,.
\end{equation}
One can see that the region $q^2 - q_{\bot}^2 \sim (\hbar\delta)^2 $ gives
the main contribution (for $b\delta \ll 1$). Thus, we can replace the value $q_\bot$
by $q$ in Eqs. ({38}),({39}), and then we can also use Eqs. ({42}),({43}).

Fig. 2 illustrates the structure factors for an ideal (10,10) single wall nanotube
superlattice, and for this lattice with the random distribution of angle shifts.
One can see that the values  of the structure factors are smaller, in the latter
case, at large enough $q$-quantities. However, the first few  factors
are the same, in between. 

Fig. 3 illustrates the behavior of the
$\langle \langle{S}({\bf{q}}){S^{*}}({\bf{q}})\rangle \rangle$
value  as a function of the transferred
momentum. The thin curve describes this function, according to Eq. ({75}), and the thick one
describes this, as a result of averaging over thermal fluctuations (see also Eq. ({49})).
The smooth curve of  middle thickness represents the behaviour of the function
$N(1-\exp{(-A\gm^2)})$. One can see that all the above pointed functions
 tend to $N$, at large enough $q$-values.

Now we can calculate the incoherent cross section of the investigated  process.
For this purpose, we find the functions
$\psi^a_{1s}$and $\psi^a_{2s}$ (see Fig. 4).
 One can see
that these functions are slightly smaller than for an amorphous medium.
It should be noted that the calculations were carried out for a three-dimensional
structure of nanotubes.
Our estimate shows that, for a two-dimensional nanotube lattice,
the incoherent contribution is practically the same as in an amorphous
medium.

The differential intensity of the coherent bremsstrahlung is shown in Fig. 5.

\subsection{Scheelite structures}
In this section we consider the coherent bremsstrahlung in three-atomic
single crystals of the scheelite type. For specific calculations, we select
$PbWO_4$ and $CaWO_4$ single crystals. It is interesting to notice that
$PbWO_4$ single crystals are widely used for the realization of
electromagnetic calorimeters \cite{Brennan:2002jj}. In Ref. \cite{Baskov:1999md} it was shown that
the coherent radiation in such structures influences some characteristics of the calorimeters.

A crystallographic structure of the scheelite type is shown in
Fig. 6.
The fundamental cell is represented by a the tetragonal prism with
the side of the squared basis and the height which are equal to
5.44 (5.22)
and 12.01 (11.45) angstroms for the $PbWO_4$ ($CaWO_4$)
single crystal, respectively. The fundamental cell contains
4 lead (calcium), 4 tungsten and 16 oxygen atoms. The oxygen
atoms are located at the corners of the tetragons around the
tungsten atoms.

Using relations similar to Eqs. ({55})-({61}) we have calculated the differential
intensity ($x d\sigma_{CB}/dx$) and the linear polarization
of coherent bremsstrahlung in $PbWO_4$ and $CaWO_4$ single crystals.
In these calculations we use three different amplitudes
of thermal fluctuations. One can expect that the energy
of the fluctuations is the same for every sort of atoms \cite{Kit}.
This means that the amplitude of fluctuations is inversely proportional
to the atomic mass. For tungsten atoms we select an amplitude equal
to 0.04 angstroms (as in the tungsten single crystal at room temperature).
Note that the intensity of the coherent radiation depends weakly enough
on the amplitude of fluctuations, and hence our approach is justified.

 The calculations were carried out for case when the electron
momentum lies in the {\it yz}-plane (see Figs. 7,8) and the angle between the
z-axis and the direction of motion is equal to 5 mrad. One can see that
the behavior of the curves is different for both single
crystals. Note that the difference in the lattice constants of
both single crystals is small and, because of this, it cannot
yield the explanation of the effect. For an understanding of the effect,
we should  take into account that in monoatomic structures
the values of the structure factors determine the set of allowed
points of the reciprocal lattice. In our case, the
$\langle Y\rangle \langle Y^*\rangle $-values play an analogous role.
The different behavior of these values
for the two different single crystals is the main issue in
the description of coherent processes.
So, the coherent bremsstrahlung in a $PbWO_4$ single crystal is
similar to this process in a monoatomic single
crystal. Indeed, the contribution in the intensity of oxygen
atoms is small, due to the low Z-value. For the tungsten and lead atoms,
the form factors and charges are approximately equal in between,
and the allowed points are approximately defined with the help of
the factor $\sum_{j=1}^8 \exp({-{\bf{g}} {\bf{r}})_j})$.

The degree of linear polarization can be defined from the relations

\begin{equation}
\label{77}
P(x)={2(1-x) \alpha_{QED} r_e^2 \psi_3(\delta) \over x d\sigma(x)/dx}\,,
\end{equation}
\begin{equation}
\label{78}
\psi_3(\delta)={4(2\pi)^2 \delta^3 \over V} \sum_{{\bf{g}}}
\langle Y \rangle \langle Y^* \rangle
{(g_x^2-g_y^2)\cos(2\beta) + 2g_xg_y \sin(2\beta) \over g^4 g_{||}^4} \,,
\end{equation}
where $g_x, g_y$ are the components of the ${\bf{g}}$-vector on the
$x$ and $y$-axes and $\beta$ is the angle between the $yz$-plane and the
arbitrary plane where the $z$-axis is located \cite{Tim}.
Figs. 7,8 illustrate also the degree of  linear polarization.
The maximal polarization takes place when in Eq. ({77})
the incoherent contribution is significantly less than the coherent one.
This situation may be realized for high energy values of the electron beam.
\subsection{Some remarks}
It is necessary to point out the specific peculiarities of coherent
bremsstrahlung, which apply to our examples.
The intensities of coherent radiation obtained in this paper
were calculated for special orientations of the atomic structures,
and they represent so-called ideal spectra \cite{Tim}. Due to the specific
motion of electrons in the atomic structures, for the description of the real
spectra we need to take into account the additional intensity of radiation
 arising from the nearby directions of motion. This additional intensity
takes place mainly for small $x$-values. However, the above mentioned orientations
are convenient for comparison, in various conditions of the radiation
sources (such as different orientations or atomic structures), and they
are widely covered in the literature.
It should be noted that this remark does not indicate a violation of the
coherent mechanism. The reason is that very small deviations of
the direction of the electron motion from the $yz$-plane (at the orientations pointed out above)
give a sizeable contribution in the radiation intensity (at low $x$).

The intensity spectra are presented for one electron energy. It is easy
to understand their behavior at different energies: the incoherent
contribution is practically independent of the electron energy;
the coherent intensity is proportional to this energy, at the condition
that the orientation angle is changed in a way inversely proportional to
the energy.

The theory of coherent bremsstrahlung is violated at some orientations
of atomic structures (at high enough electron energies).
This problem may be investigated according to Ref. \cite{BKS}.

It should be noted that the calculated spectra of
coherent bremsstrahlung may be interesting for applications of this
effect. Despite the large volume of the fundamental cell,
the radiation intensity is large enough, for rather large values
of the orientation angle.

Calculations show that  the incoherent contribution  in periodic atomic
structures  is insignificantly smaller than for the corresponding amorphous ones.
However, a precise determination of the incoherent cross section may be
useful, for electron energies smaller than some GeV. In this case, the emission
angle of the $\gamma$-quantum is detectable, and the investigation of
the incoherent contribution may give additional information.

Our examples illustrate different types of probability density
functions. For a nanotube lattice this function depends only on
one coordinate, and the other coordinates are functions of the latter.
The remaining examples illustrate the case when all atoms are independent.

\section{Conclusion}
In this paper we have shown that, in imperfect periodic atomic structures,
the coherent part of the cross section is defined by
the averaged potential of the structure, and the incoherent (diffusion) one
is defined by the pair correlation functions. The method considered here
allows one to solve the problem, on the basis of the normalized probability
density function. In particular, we have calculated the cross section
of coherent  bremsstrahlung in polyatomic single crystals
with different thermal fluctuations amplitudes.
We also considered further developments of our method for atomic structures
with a variable number of atoms in the cell, fluctuating periods, etc.

In this paper we did not investigate the process of coherent $e^\pm$-pair production
in periodic structures. However, in this case there is no problem in writing similar equations
as for coherent bremsstrahlung, using well known equations for the process
and the relations obtained here.

On the basis of our considerations, we think that coherent bremsstrahlung
and $e^{\pm}$-pair production, at particle energies
between a few hundreds of MeV and some GeV, may be utilized for
the investigation and characterization of the atomic structures,
in parallel with other methods, such as x-ray diffraction.
At these energies,
the collimation of the electron beam (or the measurement of the photon angle emission)
is possible, and therefore the possibility of a detailed study
of pair correlations appears.

In whole, our investigations may be useful for the search of new sources
of coherent radiation. The examples considered in the paper
are illustrative of such a possibility.

\newpage
\section{Figure captions}
Fig. 1.  The three-dimensional superlattice of (10,10) armchair single-wall nanotubes and
Cartesian coordinate system ($xyz$). The fundamental cell of the structure is
presented with the help of the thick lines. The black points are atoms of the
nanotubes. OA = AB = BC = CO = 17 \AA, b = 2.4 \AA, radius of
circles R= 6.8 \, \AA.

Fig. 2. Structure factors for the superlattice of (10,10) armchair single-wall nanotubes
as a function of the transferred momentum $q$ (in mc units). The circles
present the ideal structure and the black points present the structure
with the random distribution of the angle shifts. The curve is
the function $N_a^2 J_0^2(qR)$.

Fig. 3. The behavior of
$(\langle \langle{S}({\bf{q}}){S^{*}}({\bf{q}}) \rangle \rangle$
  as a function of the transfered
momentum in $mc$-units (see explanations in the text).

Fig. 4. Dependence of the functions $\psi_{1BG}^{a},
 \psi_{2BG}^{a}$ (1,2) and $\psi_{1s}^{a}, \psi_{2s}^{a}$ 
on the minimal transfer momentum (in $mc$-units). 
The curves $1^{'},\, 2^{'}$
represent the result of averaging over thermal fluctuations,
and the curves $1^{"},\,2^{"}$ represent the total result of
averaging over both thermal fluctuations and angle shifts.

Fig. 5. Differential photon spectra in  the (10,10) armchair single wall nanotube superlattice
for the structure  at fixed angle $\varphi_{1,1}=0$ (1)
 and for one with random angle shift distribution (2).
The electron energy is equal to 10 GeV. The electron beam moves in the ($xz$) plane
(see Fig. 1) under 0.018 radian with respect to the $z$-axis.
Curves 3,4,5 illustrate the incoherent contribution for conditions
corresponding to the cases $(1,2),\, (1^{'}, 2^{'})\,(1^{"},2^{"})$ (see Fig. 4),
respectively.

Fig. 6. Three projections of atoms in the fundamental cell of
the  scheelite  crystallographic structure.

Fig. 7. a) Differential photon spectra in a $PbWO_4$ single crystal (1),
in equivalent amorphous medium (2) and
incoherent contribution (3) in the intensity as functions of
the relative photon energy.
b) Degree of the linear polarization (1) and its maximum value
(2).The electron energy is equal to 10 GeV. The electron beam moves in the ($yz$) plane
(see Fig. 6) under 0.005 radian with respect to the $z$-axis.

Fig. 8. The same as in Fig. 7 but for a $CaWO_4$ single crystal.

\newpage
\begin{figure}
\scalebox{0.7}{\includegraphics{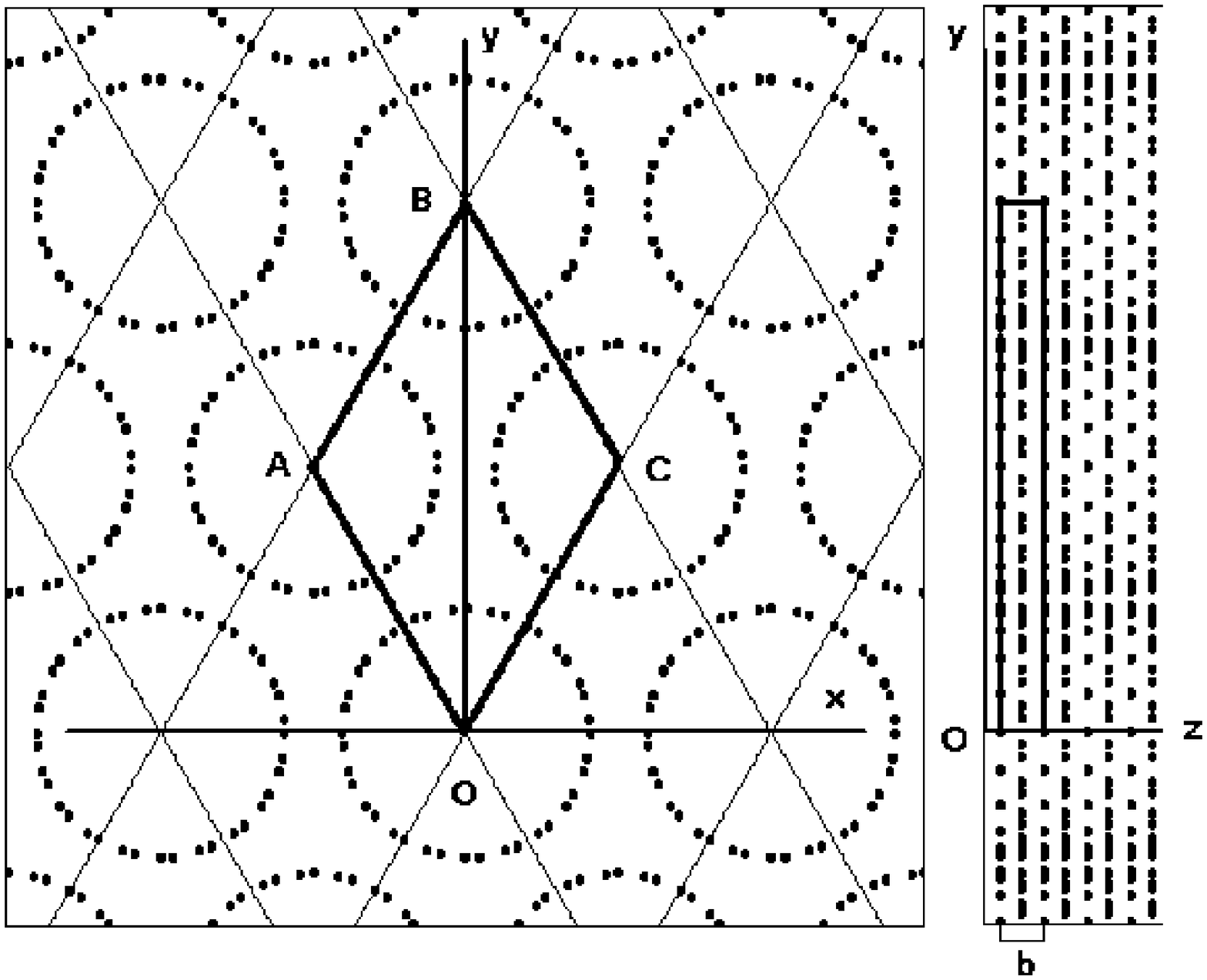}}
{\caption{
              }}
\end{figure}

\newpage
\begin{figure}
\scalebox{0.7}{\includegraphics{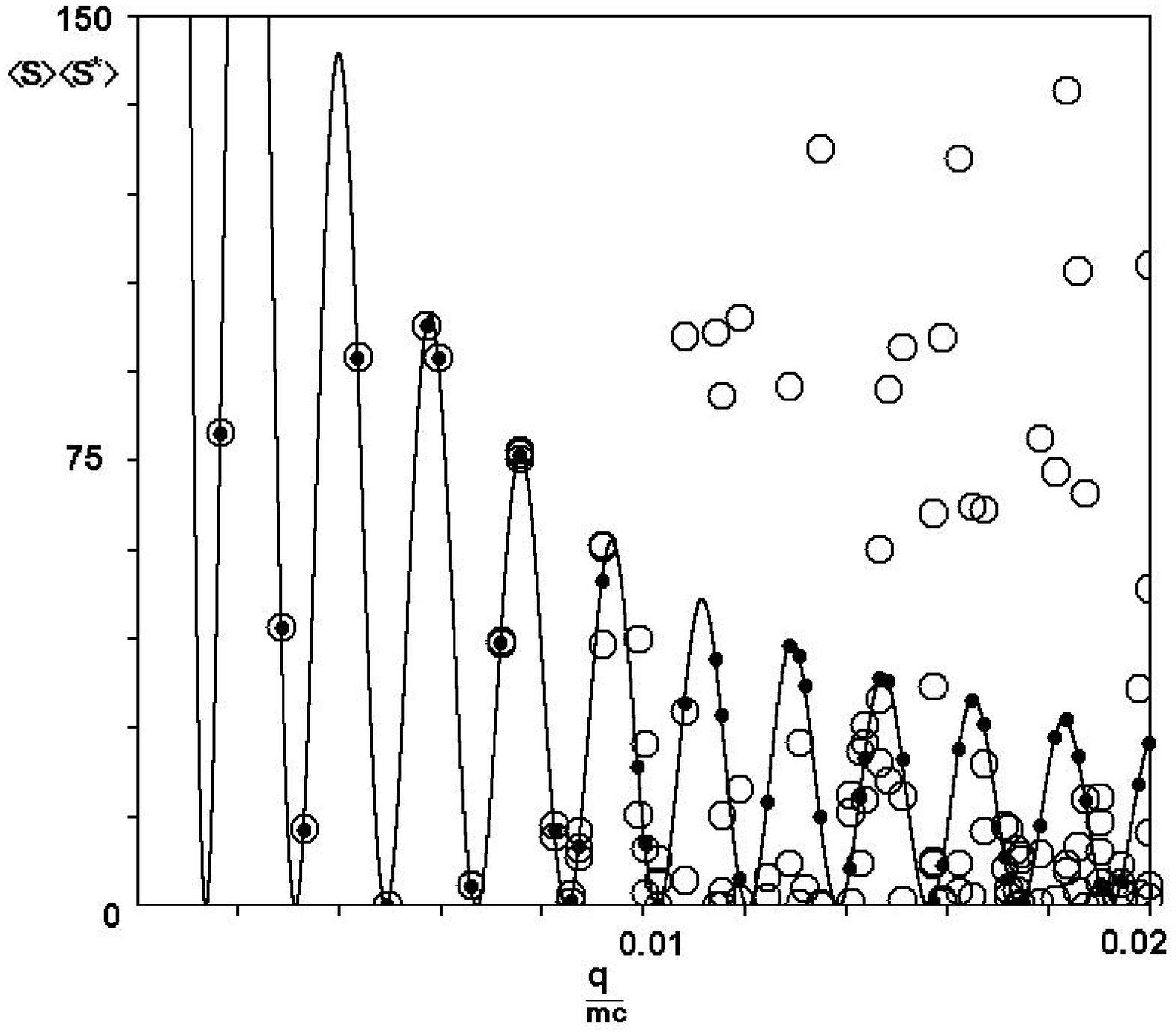}}
{\caption{
              }}
\end{figure}
\vspace*{3 cm}

\newpage
\begin{figure}
\scalebox{0.7}{\includegraphics{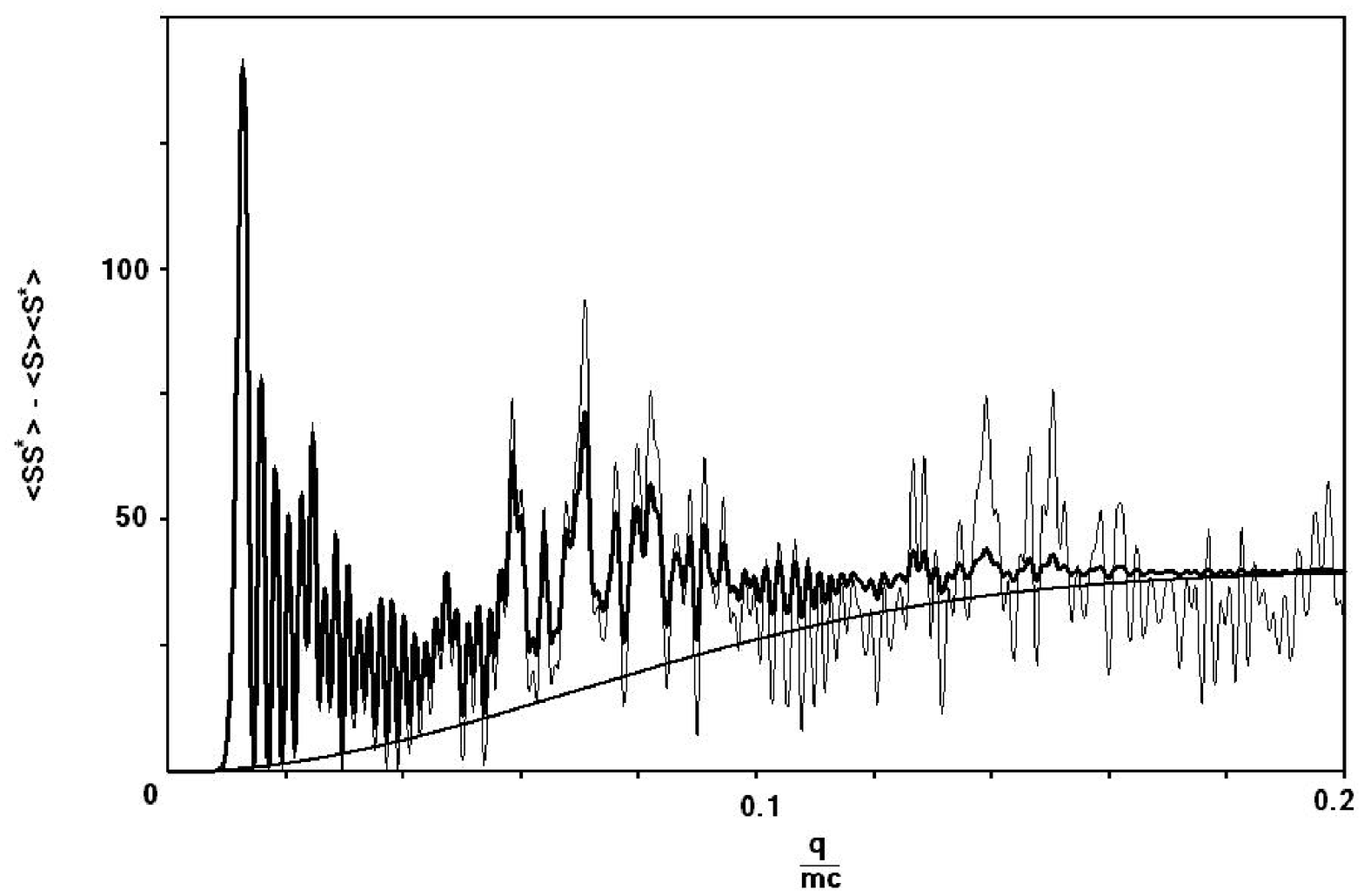}}
{\caption{
              }}
\end{figure}
\vspace*{10 cm}

\newpage
\begin{figure}
\scalebox{0.7}{\includegraphics{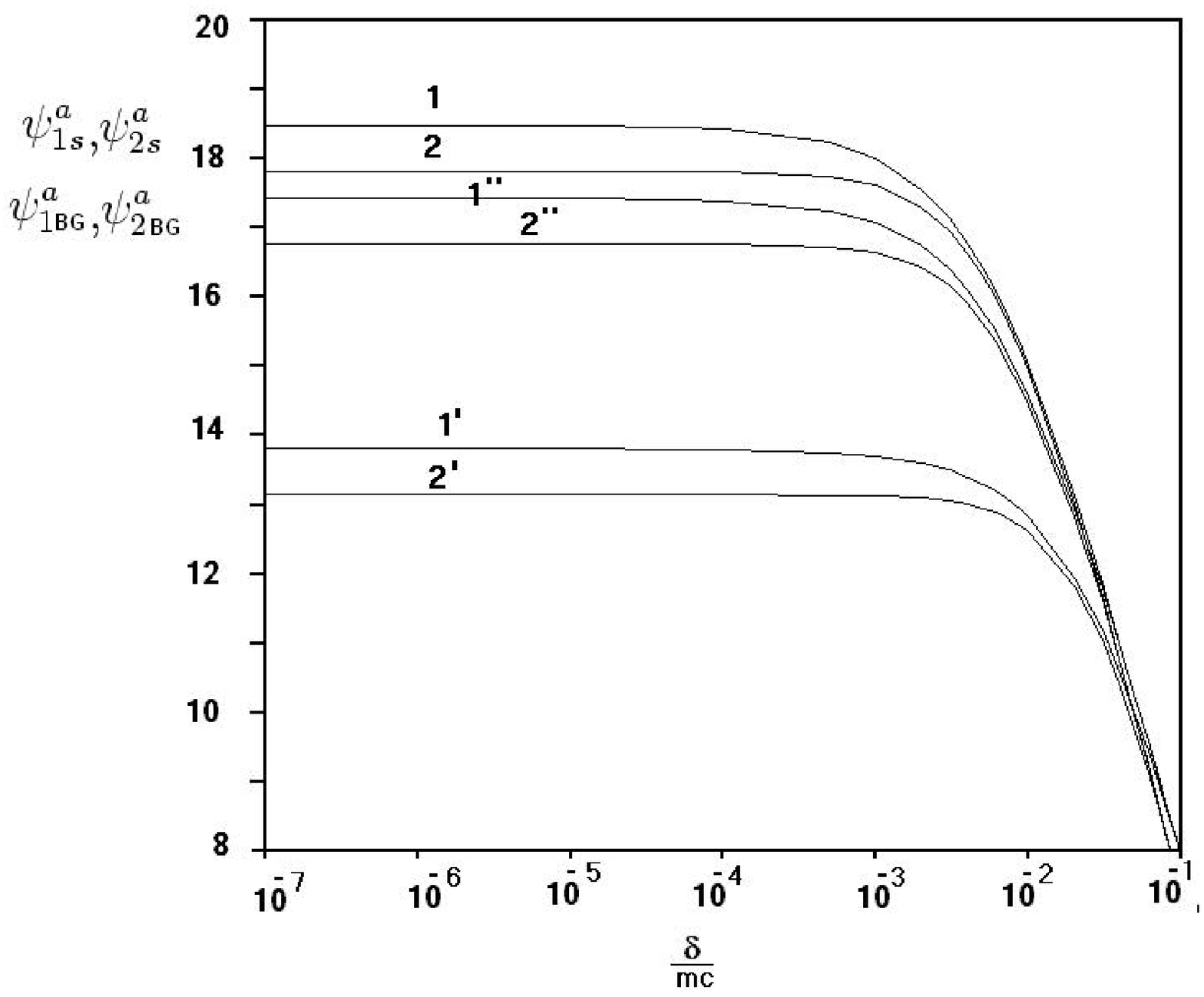}}
{\caption{
       }}
\end{figure}

\newpage
\begin{figure}
\scalebox{0.7}{\includegraphics{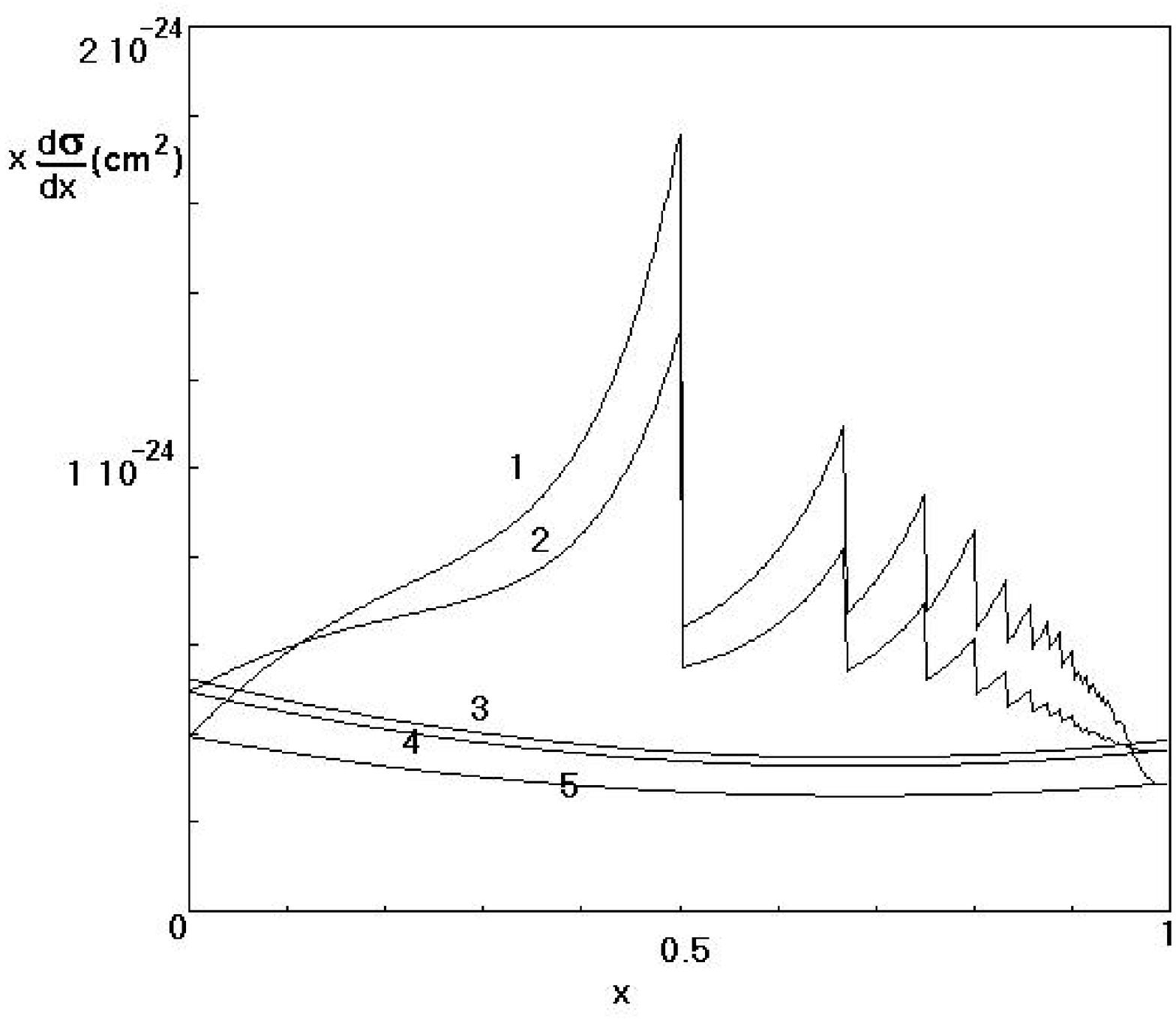}}
{\caption{
              }}
\end{figure}

\newpage
\begin{figure}
\scalebox{0.7}{\includegraphics{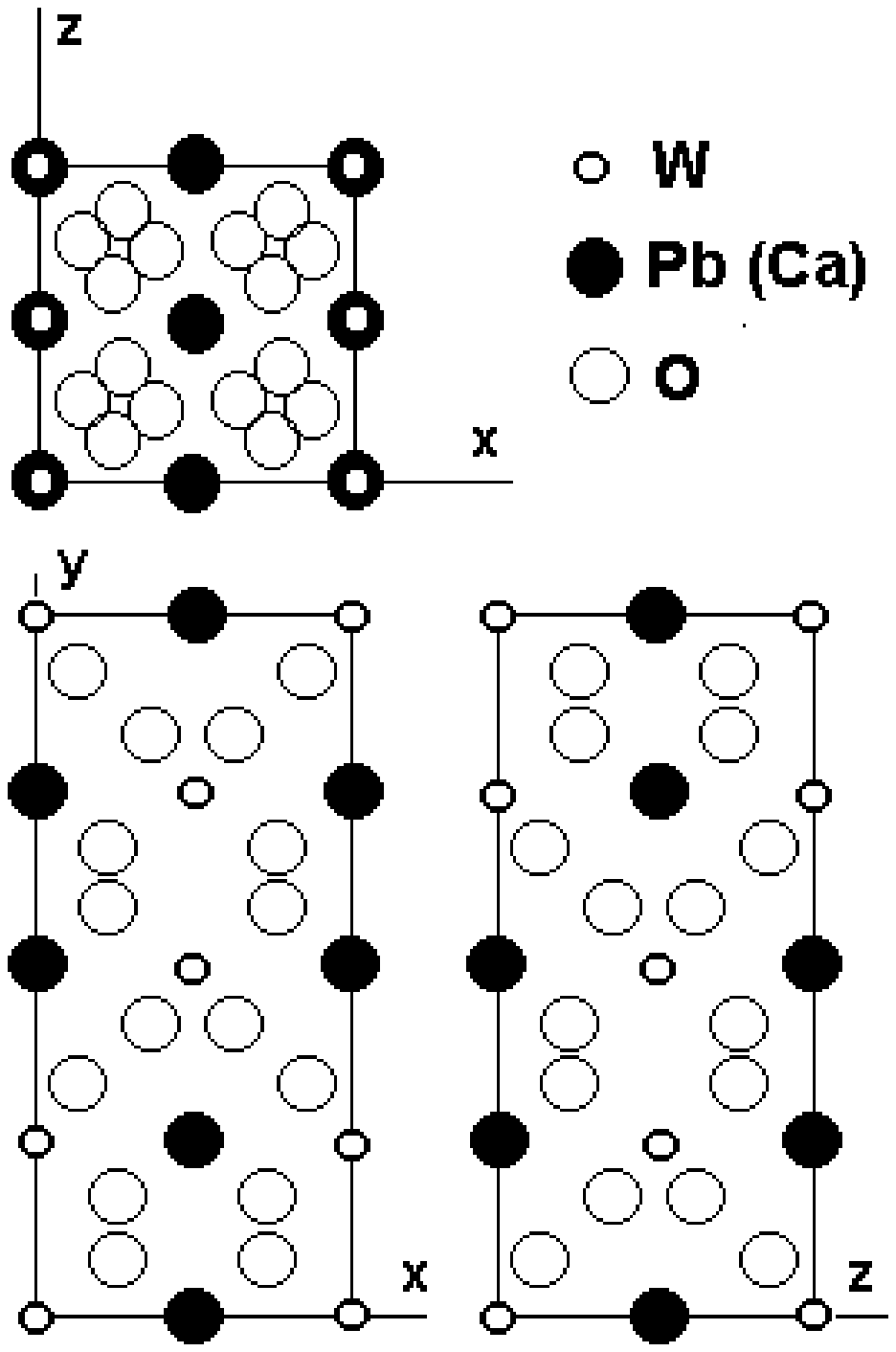}}
{\caption{
              }}
\end{figure}
\newpage
\begin{figure}
\scalebox{0.7}{\includegraphics{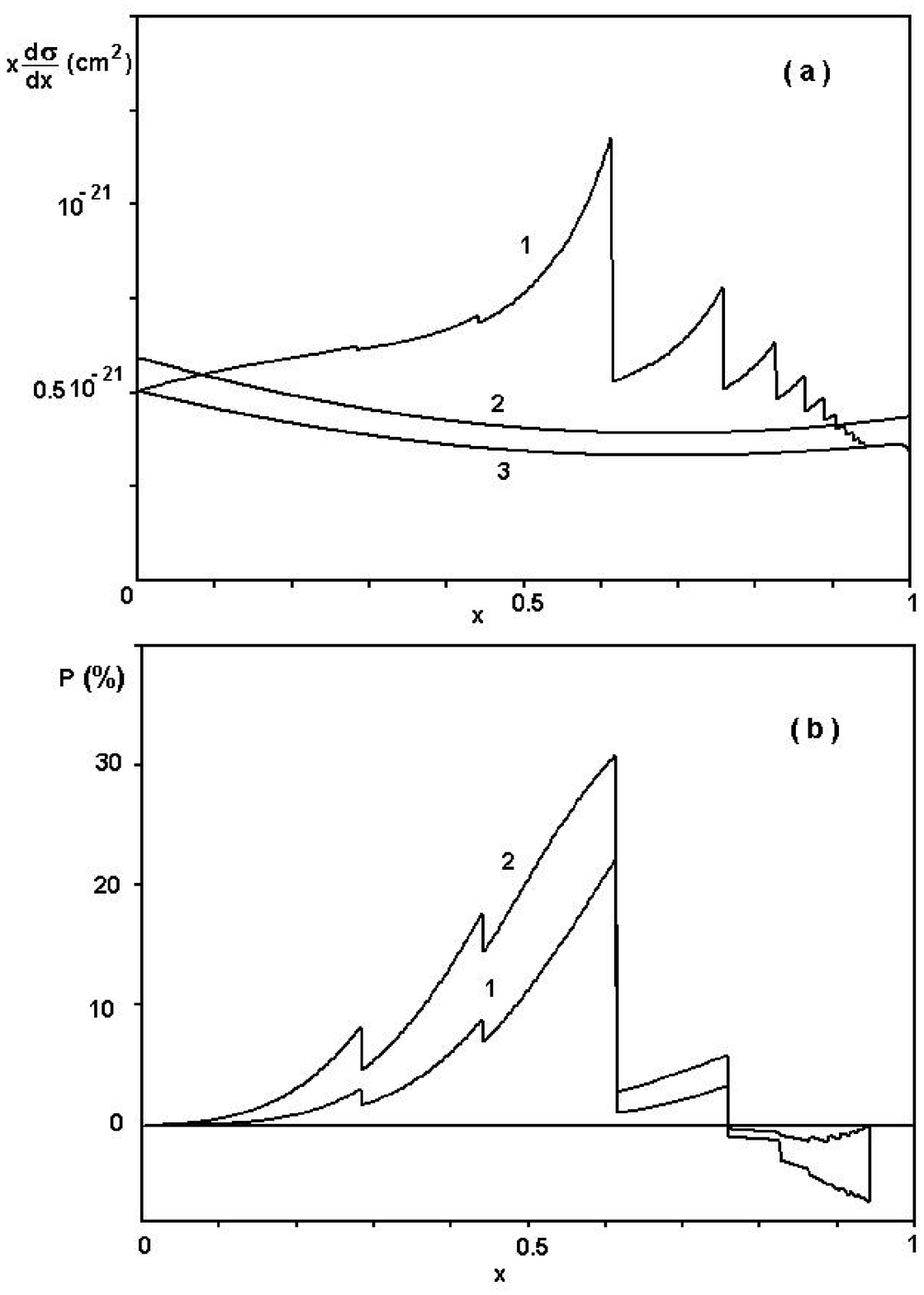}}
{\caption{
              }}
\end{figure}

\newpage
\begin{figure}
\scalebox{0.7}{\includegraphics{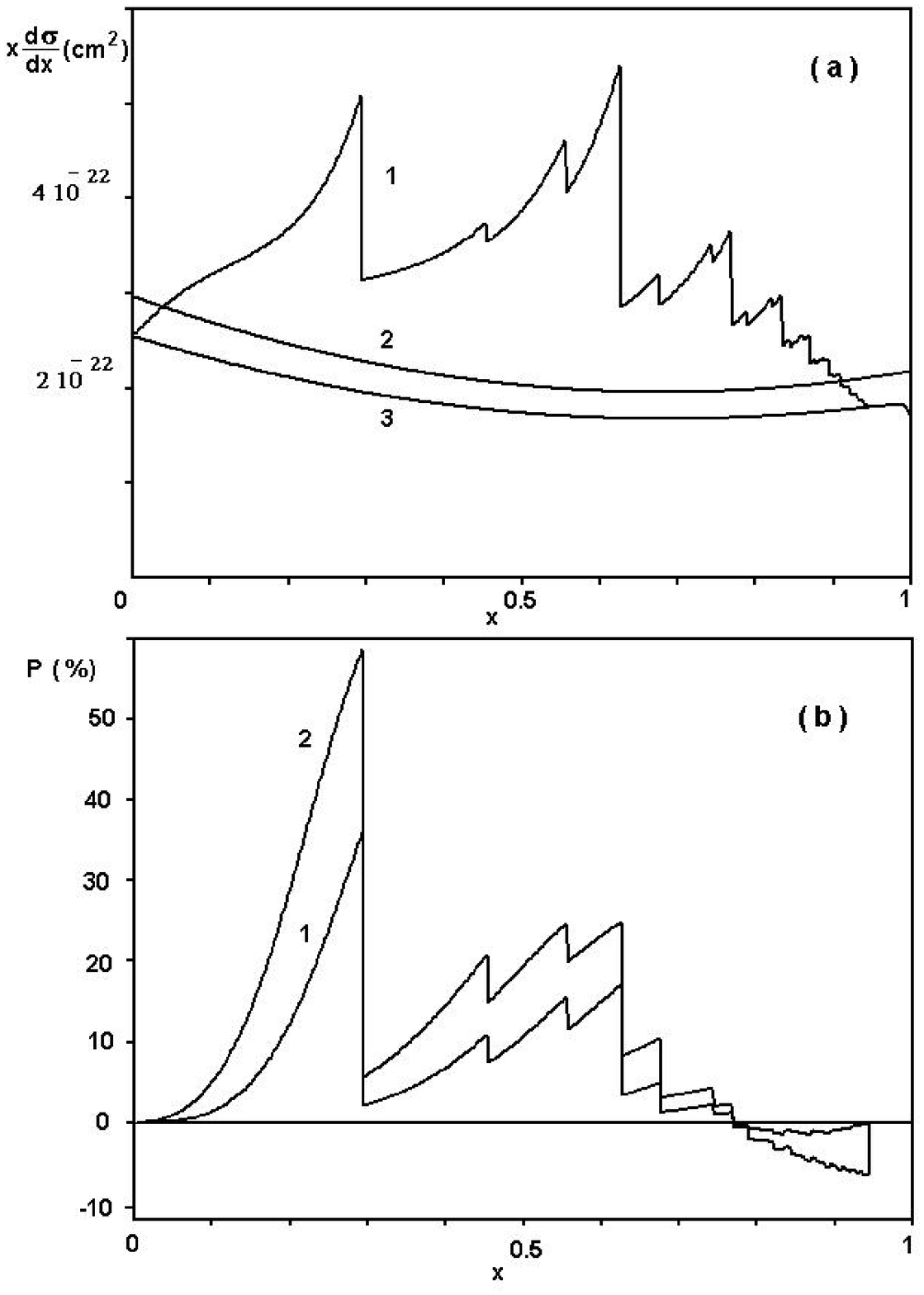}}
{\caption{
              }}
\end{figure}
\end{document}